\documentclass[aps,prstab,superscriptaddress, twocolumn]{revtex4-2}
\usepackage{hyperref}
\usepackage{graphicx}   %% Sample images
\usepackage{setspace}
\usepackage{enumitem}

\begin{document}

%\preprint{}

\title{Production of neptunium and plutonium nuclides from uranium carbide using 1.4-GeV protons}


\author{M.~Au}
\email[]{Corresponding author: mia.au@cern.ch}
%\thanks{}
\affiliation{European Organization for Nuclear Research (CERN), Meyrin, 1211 Geneva, Switzerland}
\affiliation{Johannes Gutenberg-Universit{\"a}t Mainz, 55099 Mainz, Germany}
\author{M.~Athanasakis-Kaklamanakis}
\affiliation{European Organization for Nuclear Research (CERN), Meyrin, 1211 Geneva, Switzerland}
\affiliation{Katholieke Universiteit Leuven, B-3001 Leuven, Belgium}
\author{L.~Nies}
\affiliation{European Organization for Nuclear Research (CERN), Meyrin, 1211 Geneva, Switzerland}
\affiliation{University of Greifswald, 17489 Greifswald, Germany}
\author{R.~Heinke}
\affiliation{European Organization for Nuclear Research (CERN), Meyrin, 1211 Geneva, Switzerland}
\author{K.~Chrysalidis}
\affiliation{European Organization for Nuclear Research (CERN), Meyrin, 1211 Geneva, Switzerland}
\author{U.~K{\"o}ster}
\affiliation{European Organization for Nuclear Research (CERN), Meyrin, 1211 Geneva, Switzerland}
\affiliation{Institut Laue-Langevin, 38000 Grenoble, France}
\author{P.~Kunz}
\affiliation{TRIUMF, V6T 2A3 Vancouver, Canada}
\author{B.~Marsh}
\affiliation{European Organization for Nuclear Research (CERN), Meyrin, 1211 Geneva, Switzerland}
\author{M.~Mougeot}
\altaffiliation[]{Present address: Univeristy of Jyv{\"a}skyl{\"a}, 40014 Jyv{\"a}skyl{\"a}, Finland}
\affiliation{European Organization for Nuclear Research (CERN), Meyrin, 1211 Geneva, Switzerland}
\affiliation{Max Planck Institut f{\"u}r Kernphysik, 69117 Heidelberg, Germany}
\author{L.~Schweikhard}
\affiliation{University of Greifswald, 17489 Greifswald, Germany}
\author{S.~Stegemann}
\affiliation{European Organization for Nuclear Research (CERN), Meyrin, 1211 Geneva, Switzerland}
\author{Y.~Vila Gracia}
\affiliation{European Organization for Nuclear Research (CERN), Meyrin, 1211 Geneva, Switzerland}
\author{Ch.~E.~D{\"u}llmann}
\affiliation{Johannes Gutenberg-Universit{\"a}t Mainz, 55099 Mainz, Germany}
\affiliation{GSI Helmholtzzentrum f{\"u}r Schwerionenforschung, 64291 Darmstadt, Germany}
\affiliation{Helmholtz Institute Mainz, 55099 Mainz, Germany}
\author{S.~Rothe}
\affiliation{European Organization for Nuclear Research (CERN), Meyrin, 1211 Geneva, Switzerland}


\date{\today}

\begin{abstract}
Accelerator-based techniques are one of the leading ways to produce radioactive nuclei. In this work, the Isotope Separation On-Line method was employed at the CERN-ISOLDE facility to produce neptunium and plutonium from a uranium carbide target material using 1.4-GeV protons. Neptunium and plutonium were laser-ionized and extracted as 30-keV ion beams. A Multi-Reflection Time-of-Flight mass spectrometer was used for ion identification by means of time-of-flight measurements as well as for isobaric separation. Isotope shifts were investigated for the 395.6-nm ground state transition in $^{236,237,239}$Np and the 413.4-nm ground state transition in $^{236,239,240}$Pu. Rates of $^{235-241}$Np and $^{234-241}$Pu ions were measured and compared with predictions of in-target production mechanisms simulated with G\textsc{eant}4 and FLUKA to elucidate the processes by which these nuclei, which contain more protons than the target nucleus, are formed. $^{241}$Pu is the heaviest nuclide produced and identified at a proton-accelerator-driven facility to date. We report the availability of neptunium and plutonium as two additional elements at CERN-ISOLDE and discuss the limit of accelerator-based isotope production at high-energy proton accelerator facilities for nuclides in the actinide region. 
\end{abstract}

%\keywords{}

\maketitle

\section{Introduction}\label{intro}
The actinide region of the nuclear chart is a focus of research for topics including the r-process and astrophysical isotopic abundances \cite{Holmbeck2019}, nuclear fission \cite{Mumpower2018,Andreyev2018}, nuclear medicine \cite{Birnbaum2018,Robertson2018}, environmental monitoring \cite{Kudo2001}, and energy production \cite{Lee2020}. Experimental measurements of nuclear masses, $\beta$-decay, neutron capture, and fission properties are required to benchmark theoretical nuclear structure models, but are missing for many nuclei \cite{Kajino2017,Schmidt2018}. Laser spectroscopy has the capability to reveal both nuclear and atomic information through techniques such as high-resolution resonance ionization and collinear laser spectroscopy \cite{Sonnenschein2015a,Voss2017,Block2021}. Resonance ionization laser schemes have been developed for many actinide elements \cite{Kunz2004,Kneip2020}, facilitating the use of laser ionization as a spectroscopic technique, a tool for efficient and element-selective production of radioactive ion beams, as well as for trace element detection \cite{Raeder2019}.

All actinide nuclides are radioactive. $^{232}$Th and $^{235,238}$U are available naturally in macroscopic quantities, but the majority of the other actinide isotopes must be artificially produced, with production often being one of the most limiting factors in their study and use. Research reactors are the leading method of production for the study of some actinide isotopes, producing a selection of species \cite{Roberto2015}. Some specific isotopes can also be obtained through generators made from decaying parent nuclides, for example $^{227}$Th (and $^{227}$Ra) produced from $^{227}$Ac \cite{McAlister2011}. Alternatively, actinides can be produced at accelerator facilities using a projectile on a thin target to induce fusion-evaporation reactions or multi-nucleon transfer. The reaction products can then be separated in-flight as ion beams; long-lived isotopes can be collected for off-line isolation and use \cite{Blumenfeld2013}. The Isotope Separation On-Line (ISOL) method uses an energetic driver beam that interacts with a thick target, generating reaction products that diffuse through the target material and effuse to the ion source where they are ionized and extracted as an ion beam. The ISOLDE facility at CERN uses 1.4-GeV protons from the CERN Proton Synchrotron Booster to produce and deliver more than 1000 different nuclide species \cite{Catherall2017,Ballof2020}. Uranium carbide (UC$_\textrm{x}$) from depleted uranium is one of the most commonly used target materials \cite{Ramos2020,Gottberg2016}. The ISOL method is typically able to provide isotopes lighter than the target nucleus through fission, fragmentation, and spallation reactions. The production of above-target elements (higher proton number $Z$ than the target nucleus) has barely been studied. In the case of the $^{238}$U target nucleus, above-target production pathways would reach into the elusive transuranium region. 
\begin{figure*}
	\centering
	\includegraphics[width=0.8\textwidth]{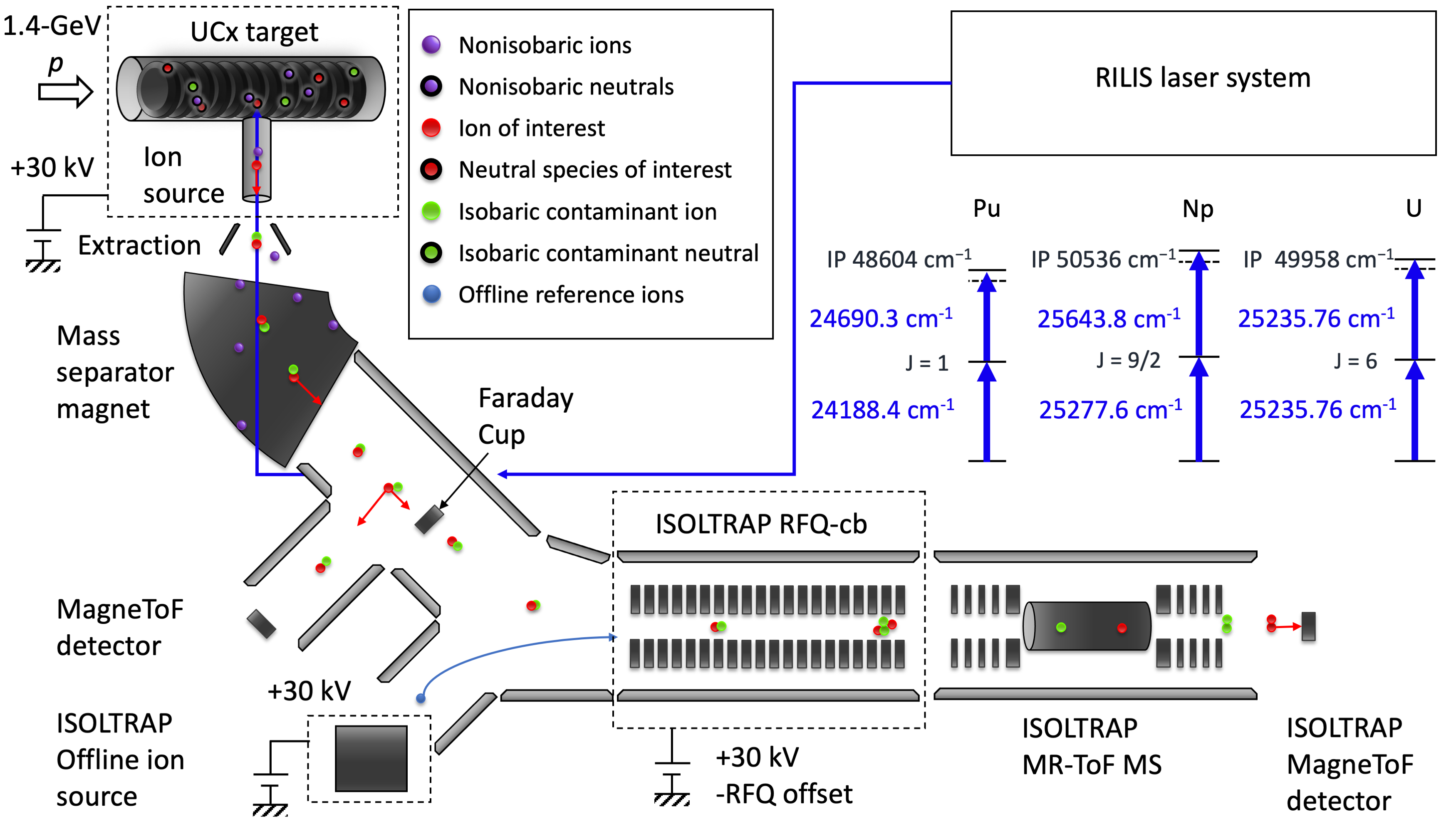}%
	\caption{Schematic of the experimental setup. Isotopes are generated in the UC$_\textrm{x}$ target by 1.4-GeV protons (up to 2\,$\mu$A) from the CERN Proton Synchrotron Booster. Reaction products diffuse and effuse into the hot cavity ion source where they can be resonantly ionized by lasers from RILIS \cite{Fedosseev2017}. Ions are extracted as a beam via the 30-kV extraction electrode and separated by their mass-to-charge ratio in the separator magnet. The mass-separated beam is either sent to a MagneToF detector \cite{ETPmagnetof} or through the central beamline to ISOLTRAP, where the ions are cooled and bunched in the RFQ-cb and sent to the MR-ToF MS for isobaric mass separation. \label{fig:schematic}}
\end{figure*}

Beams of $^{238}$Np (2019) and $^{239}$Pu (2013) were produced at TRIUMF, Canada from the ISOL method with 500-MeV protons from two separate UC$_\textrm{x}$ targets using resonance laser ionization in rhenium hot cavity surface ion sources, and isotopic yields were reported \cite{TRIUMFyields, Kunz2023}. The extraction behaviour of actinide beams from the UC$_\textrm{x}$ target matrix still remains largely uncharacterized. Experimental data on availability, intensity and purity of Np and Pu ion beams will enable experiments on these elements at ISOL facilities. Identifying the high-$Z$ limits of the accelerator-based production technique at proton accelerator facilities gives essential information for the field of actinide research. Theoretical estimates can be provided by models describing the processes occurring in collisions of 1.4-GeV protons with target nuclei like the Monte Carlo code FLUKA \cite{FLUKA1,FLUKA2}, which is used to describe nuclide production at ISOLDE \cite{Ballof2020}.

\section{Methods}\label{methods}
The depleted UC$_\textrm{x}$ target (UC$_2$-C$_2$ with 0.25 wt\% $^{235}$U, 82.95 wt\% $^{238}$U, 16.8 wt\% C, density 3.67 g/cm$^3$) used for the experiment was irradiated using 5($10^{18}$) protons over a time of 313\,h ($\approx$2 weeks) while actively used for isotope production and extraction during an experiment ending on the 12$^\textrm{th}$ of November 2018. After a 929(1) day ($\approx$2.5 year) cooling period, first investigations were conducted using the remaining long-lived inventory. At 934.0(5) days of cooling time, another irradiation of 8.5($10^{17}$) 1.4-GeV protons was performed on the target unit. Subsequent measurements were taken both with and without direct irradiation over an experimental period ending at 942.4(5) days after the end of the initial irradiation. In this study, measurements were taken under three conditions: 
\begin{enumerate}[leftmargin=2cm]
	\item[``before, off":] {Before any additional irradiation (after the cooling period).}
	\item[``after, off":] {After re-irradiation, without the proton beam actively hitting the target.}
	\item[``after, on":] {After re-irradiation, with the proton beam actively hitting the target.}
\end{enumerate}

The FLUKA model of in-target production was used to predict production of elements up to plutonium for a beam of 1.4-GeV protons, Gaussian full-width at half-maximum (FWHM) 6.35\,mm, on a depleted UC$_\textrm{x}$ target \cite{Ballof2020}. Particle fluence spectra caused by the 1.4-GeV proton beam are modeled along with the inventory of radionuclides produced during irradiation. The simulated fluence was additionally used as input for the ActiWiz software \cite{Vincke2014} to calculate in-target inventory of long-lived actinides after 734\,days ($\approx$2 years) of cooling time \cite{Duchemin2020}. A G\textsc{eant}4 \cite{ALLISON2016geant4, G4physicsguide} package-based model developed at TRIUMF \cite{Garcia2017} was adapted to the ISOLDE proton beam energy and geometry to model reaction pathways. This was used to model $10^9$ 1.4-GeV primary protons irradiating a UC$_\textrm{x}$ target (83.2 wt\% U and 16.8 wt\% C) of natural uranium, consisting of $^{234}$U (0.005\,\%), $^{235}$U (0.72\,\%), and $^{238}$U (99.275\,\%). The difference caused by the $^{235}$U content is expected to be negligible. Excessive decay time was included in the G\textsc{eant}4 model to evaluate the largest feasible effect of previous irradiation and cooling time, i.e., all inventory has been allowed to decay, considering a decay time extending to timescales relevant to the scheduling of future experiments [9792 days ($\approx$27 years) after primary impact]. The FLUKA and G\textsc{eant}4 models were used to compare modeled rates with experimental rates achievable by using a previously irradiated target. Further details of the models can be found in Appendix \ref{appendix:models}.

The experimental setup is shown schematically in Fig.~\ref{fig:schematic}. Ions were formed using the Resonance Ionization Laser Ion Source (RILIS) \cite{Fedosseev2017,Rothe2016} and a rhenium hot cavity surface ion source. Ions were extracted as 30-keV ion beams, which were mass-separated using the General Purpose Separator (GPS) dipole magnet. The two-step ionization schemes \cite{Kneip2020} shown in Fig.~\ref{fig:schematic} were chosen, using intra-cavity second harmonic generation in titanium:sapphire (Ti:Sa) lasers \cite{Sonnenschein2015,Rothe2011} for the first excitation step (FES) and second excitation step (SES) of Np and Pu laser ionization schemes. A grating Ti:Sa laser \cite{Teigelhofer2010} was also used to perform scans of the FES wavelength, with typical laser FWHM of 7\,GHz. Simultaneous application of a uranium laser ionization scheme \cite{Savina2017} during beam composition studies enabled the identification of potential contaminants such as $^{238}$U from the target material. Further details of the ion source are described in Appendix \ref{appendix:target}.

The target and ion source are heated separately by resistive (Joule) heating, such that the ion source could be at 2000\,$^{\circ}$C for surface ionization while keeping the target at temperatures estimated to be 500(200)\,$^{\circ}$C by conduction only. With the ion source heated to facilitate surface ionization of the released species, the target temperature was increased stepwise in 100-300\,$^{\circ}$C increments from 1000\,$^{\circ}$C to a maximum of 2100\,$^{\circ}$C in the ``before, off" condition. At nominal target temperature (2000\,$^{\circ}$C), additional heat deposition from the proton beam is expected to contribute less than 10$\%$ to the total target heating power. Further details regarding the target temperature are described in Appendix \ref{appendix:target}.

Mass scans with the laser ionization switched on/off were conducted using the ISOLDE GPS separator magnet in the ``after, off" condition. Because of the potential increase in surface ionization caused by additional heating from laser power deposition, mass scans with lasers ``off'' were done with the FES blocked or detuned from resonance, and the SES on resonance. Isotope shifts are on the order of 1-2 GHz per neutron number and were not corrected for during the mass scans. In this work, ``MagneToF detector" refers to the detector immediately after the mass-separator magnet, unless ``ISOLTRAP MagneToF detector" is explicitly indicated. Rates on the MagneToF detector are reported as the Faraday cup (FC) equivalent absolute ion intensity calculated from the manufacturer-provided gain curve \cite{ETPmagnetof}.

For analysis of the beam composition using mass spectrometry, the beam from the ISOLDE GPS was cooled and bunched in the ISOLTRAP Radio-Frequency Quadrupole Cooler-Buncher (RFQ-cb) \cite{HERFURTH2001254} using helium buffer gas. Ion bunches were then injected into the ISOLTRAP Multi-Reflection Time-of-Flight Mass Spectrometer (MR-ToF MS) \cite{Wolf2013} and captured between the electrostatic mirror potentials using the in-trap lift method \cite{Wienholtz2013}. Further details can be found in Appendix \ref{appendix:beam}.

{\setstretch{1.0}
\begin{table*}
	\centering
	\caption{Production mechanisms in \% of total events for selected nuclides of interest calculated using G\textsc{eant}4 QGSP\_INCLXX+ABLA with $10^9$ 1.4-GeV proton primaries. Inelastic and Decay columns give the sums over inelastic and radioactive decay processes, respectively. The larger contribution is indicated in bold. Various capture reactions are included in the model but not shown in the table. Some processes with event fractions below 1\,\% are not shown and only the dominant parent nucleus and its corresponding percentage fraction of total events are given. The number of total events is scaled from $10^9$ protons to obtain the nuclides per $\mu$C equivalent \label{tab:G4production}}
	\begin{ruledtabular}
		\begin{tabular}{| l |r r r r r r r r | r l r |l|}
			\hline
            isotope  &\textbf{Inelastic}: & \textit{p} & \textit{n} &     \textit{d} &     \textit{t} & $^3$He & $\alpha$ & ions & \textbf{Decay}: & parent & &  nuclides$/\mu$C \\
            \hline
           $^{234}$U  & 36.4 &  28.4 &   7.2 &   0.4 &     - &      - &    - &    - &  \textbf{63.6} &  $^{234}$Pa &    59.4 & 6.7$\times10^9$ \\
           $^{236}$U  & \textbf{65.5} &  52.2 &  11.8 &   0.8 &     - &    - &            - &    - &  34.3 &  $^{236}$Pa &    34.3 & 1.3$\times10^{10}$ \\
           $^{237}$U  & \textbf{70.5} &  61.7 &   7.4 &   0.9 &   0.1 &    - &            - &    - &  29.0 &  $^{237}$Pa &    29.0 & 1.9$\times10^{10}$ \\
           $^{239}$U$^a$  & 1.9 &     - &      - &   1.3 &   0.5 &     - &  0.1 &    - &     - &      - &       - & 2.2$\times10^9$ \\
           $^{240}$U  & \textbf{100.0} &    - &     - &     - &  85.7 &    - &         11.9 &  2.4 &     - &      - &       - & 2.1$\times10^6$ \\
           $^{231}$Np  & \textbf{99.0} &   93.1 &   -   &   3.5 &   0.5 &   -  & - & - & 1.0 &   $^{231}$Pu & 1.0 & 1.3$\times10^6$ \\
           $^{232}$Np  & \textbf{100.0} & 92.3 &   0.3 &   5.4 &   0.3 &  0.1 & - & - & - &   - & - & 1.3$\times10^7$ \\
           $^{233}$Np  & \textbf{99.8} &  91.1 &   0.2 &   6.1 &   0.5 &  0.1 & - & - & 0.2 &   $^{233}$Pu & 0.2 & 3.6$\times10^7$ \\
           $^{234}$Np  & \textbf{100} &  89.9 &   0.2 &   7.7 &   0.7 &  0.1 & - & - & -   &   - & - & 2.0$\times10^8$ \\
          $^{235}$Np  & \textbf{99.8} &  88.4 &   0.2 &   8.9 &   1.0 &  0.2 & - & - & 0.2 &   $^{235}$Pu & 0.2 & 3.8$\times10^8$ \\
          $^{236}$Np  & \textbf{100.0} &  86.2 &   0.1 &  11.1 &  1.4 &  0.2 & - & - &   - &      - &       - & 8.7$\times10^8$ \\
          $^{237}$Np  & 3.7 &   3.0 &     - &   0.6 &   0.1 &      - &     - &    - &  \textbf{96.3} &   $^{237}$U &    96.3 & 2.0$\times10^{10}$ \\
          $^{238}$Np  & \textbf{100.0} &  61.0 &     - &  27.4 &   8.8 &    1.1 &          0.4 &    - &     - &      - &       - & 3.9$\times10^8$ \\
          $^{239}$Np  & 3.0 &     - &     - &   2.2 &   0.6 &    0.2 &          0.1 &    - &  \textbf{97.0} &   $^{239}$U &    97.0 & 2.2$\times10^9$ \\
          $^{240}$Np  & \textbf{73.0} &     - &     - &     - &  47.8 &    5.2 &         18.8 &  1.3 &  27.0 &   $^{240}$U &    27.0 & 7.8$\times10^6$ \\
          $^{241}$Np  & \textbf{100.0} &     - &     - &     - &     - &      - &         96.3 &  3.7 &     - &   - &       - & 6.8$\times10^5$ \\
          $^{235}$Pu  & \textbf{99.2} &  62.5 &     - &     - &     - &   18.8 &         10.9 &    - &   0.8 &  $^{235}$Am &     0.8 & 8.0$\times10^5$ \\
          $^{236}$Pu  & \textbf{97.1} &  67.8 &     - &     - &     - &   13.5 &         12.5 &    - &   2.9 &  $^{236}$Np &     1.9 & 1.3$\times10^6$ \\
          $^{237}$Pu  & \textbf{99.5} &  39.7 &     - &     - &     - &   15.5 &         39.7 &  0.3 &   0.5 &      - &       - & 2.3$\times10^6$ \\
          $^{238}$Pu  & 0.9 &   0.1 &     - &     - &     - &    0.1 &          0.7 &    - &  \textbf{99.1} &  $^{238}$Np &    99.1 & 4.0$\times10^8$ \\
          $^{239}$Pu & 0.3 &     - &     - &     - &     - &      - &          0.3 &    - &  \textbf{99.7} &  $^{239}$Np &    99.7 & 2.2$\times10^9$ \\
          $^{240}$Pu & 24.0 &     - &     - &     - &     - &    2.5 &         20.9 &  0.6 &  \textbf{76.0} &  $^{240}$Np &    75.5 & 1.0$\times10^7$ \\
          $^{241}$Pu  & \textbf{53.9} &     - &     - &     - &     - &      - &         51.7 &  2.2 &  46.1 &  $^{241}$Np &    46.1 & 1.4$\times10^6$ \\
            \hline
		\end{tabular}
	\end{ruledtabular}
$^{a}$ $^{239}$U is 98.1\,\% produced by neutron capture reactions. 
\end{table*}
}

\section{Results} \label{results}
The production mechanisms extracted from the G\textsc{eant}4 model predict that some species ($^{239,240}$U, $^{236,238,241}$Np, $^{234-237}$Pu) are entirely or primarily generated by inelastic reactions. Others ($^{237,239}$Np, $^{238,239}$Pu) form primarily through decay of a parent nucleus. The reaction pathways summarized in Table \ref{tab:G4production} predict different possibilities for in-target inventory of these species using pre-irradiated targets compared to the situation during direct irradiation. 

The isotopes of Np, Pu and Am observed with at least one event in the G\textsc{eant}4 model with $10^9$ primary proton projectiles are: $^{228-244}$Np, $^{234-244}$Pu and $^{235-244}$Am. Several of these isotopes ($^{228-229,242-244}$Np, $^{242-244}$Pu, and $^{235-240, 242-244}$Am) were predicted with event rates less than 100 per $10^9$ primaries and are therefore not included in Table \ref{tab:G4production}. The G\textsc{eant}4 model suggests that the production of some isotopes features a large contribution from decay. For example, the dominant mechanism to form $^{239}$Pu from the target material is $^{238}$U undergoing neutron capture to $^{239}$U, which decays with a half-life of 23.45 min through $^{239}$Np (2.3 days) into $^{239}$Pu, with 98.2\,\% of the $^{239}$U produced from neutron capture reactions, 97.1\,\% of the $^{239}$Np produced by $^{239}$U decay, and 99.7\,\% of the $^{239}$Pu events produced by $^{239}$Np decay. This makes $^{239}$Pu, and other long-lived isotopes with significant decay feeding, available in measurable quantities without continuous proton irradiation. Other isotopes, such as $^{236,237}$Pu, are predicted to form primarily through reactions rather than decay, particularly in the case of $^{237}$Pu since there is no contribution from in-target decay of the long-lived $^{237}$Np. With the exceptions of  $^{241}$Am (162 events, 97.5\,\% from decay of $^{241}$Pu), and $^{243}$Am, (7 events, 5 from decay of $^{243}$Pu), all Am isotopes were predicted to come entirely from inelastic reactions with protons and heavy ions. These isotopes produced dominantly through direct reactions are thus not expected to be available in off-line operation. 

\begin{figure}
	\includegraphics[width=0.5\textwidth]{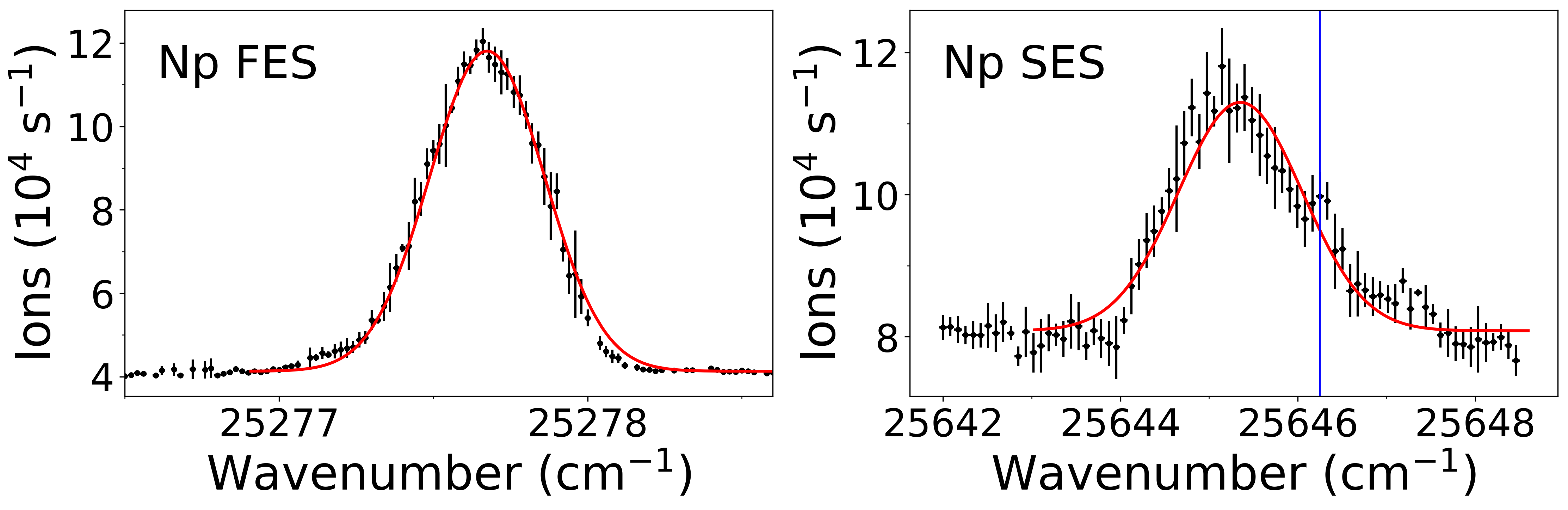}%
	\caption{Np laser ionization resonance on mass 237 at a target temperature of 2110\,$^{\circ}$C and the ``after, off" condition. Rates from the GLM MagneToF detector. Left: FES scan using 25645\,cm$^{-1}$ for the SES. The Gaussian fit shown gives a peak width of 6.3 GHz. Right: SES scan using 25277.8\,cm$^{-1}$ for the FES. The Gaussian fit shown gives a peak width of 27.57\,GHz which is typical for auto-ionizing states. Blue line: a shoulder at 25646.25\,cm$^{-1}$ which could come from another atomic or molecular ion with the same mass-to-charge ratio. \label{fig:Np_resonances}}
\end{figure}

\begin{figure}
	\includegraphics[width=0.5\textwidth]{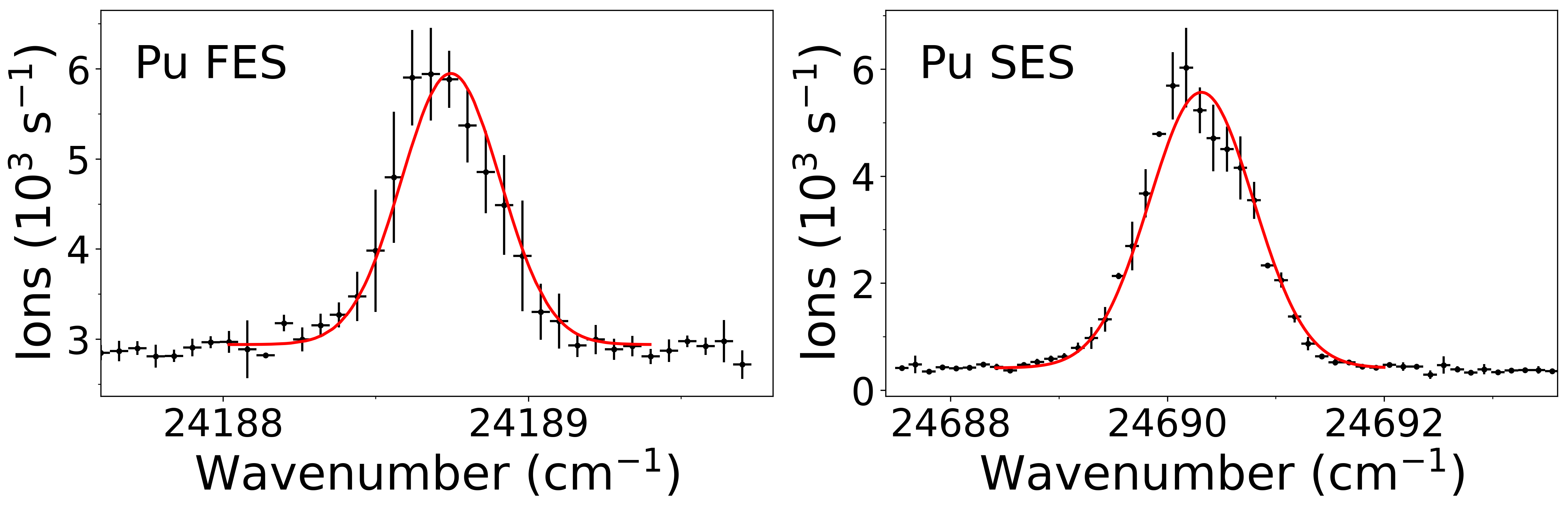}%
	\caption{Pu laser ionization resonance on mass 239 at a target temperature of 1990\,$^{\circ}$C and the ``after, off" condition. Rates from the MagneToF detector. Left: FES scan using 24690.4\,cm$^{-1}$ for the SES. Gaussian fit gives a peak width of 5.7\,GHz. Right: SES scan using 24188.4\,cm$^{-1}$ for the FES. Gaussian fit gives a peak width of 33.8\,GHz. \label{fig:Pu_resonances}}
\end{figure}

Pu ion beams were detectable at the lowest target temperatures among the studied actinide beams, with resonant laser response visible on the MagneToF detector at a target temperature of at least 1150\,$^{\circ}$C in the ``before, off" condition. Np ion beams were identified only at target temperatures above 2100\,$^{\circ}$C. The laser ionization schemes presented in Fig.~\ref{fig:schematic} for Np and Pu showed resonant enhancement on nominal mass 237 (Fig.~\ref{fig:Np_resonances}) and 239 (Fig.~\ref{fig:Pu_resonances}) beams, respectively, corresponding to the isotopes $^{237}$Np and $^{239}$Pu, which were predicted to have the largest in-target inventory for the elements Np and Pu (Tab.~\ref{tab:G4production}). 

\begin{figure}
\includegraphics[width=0.45\textwidth]{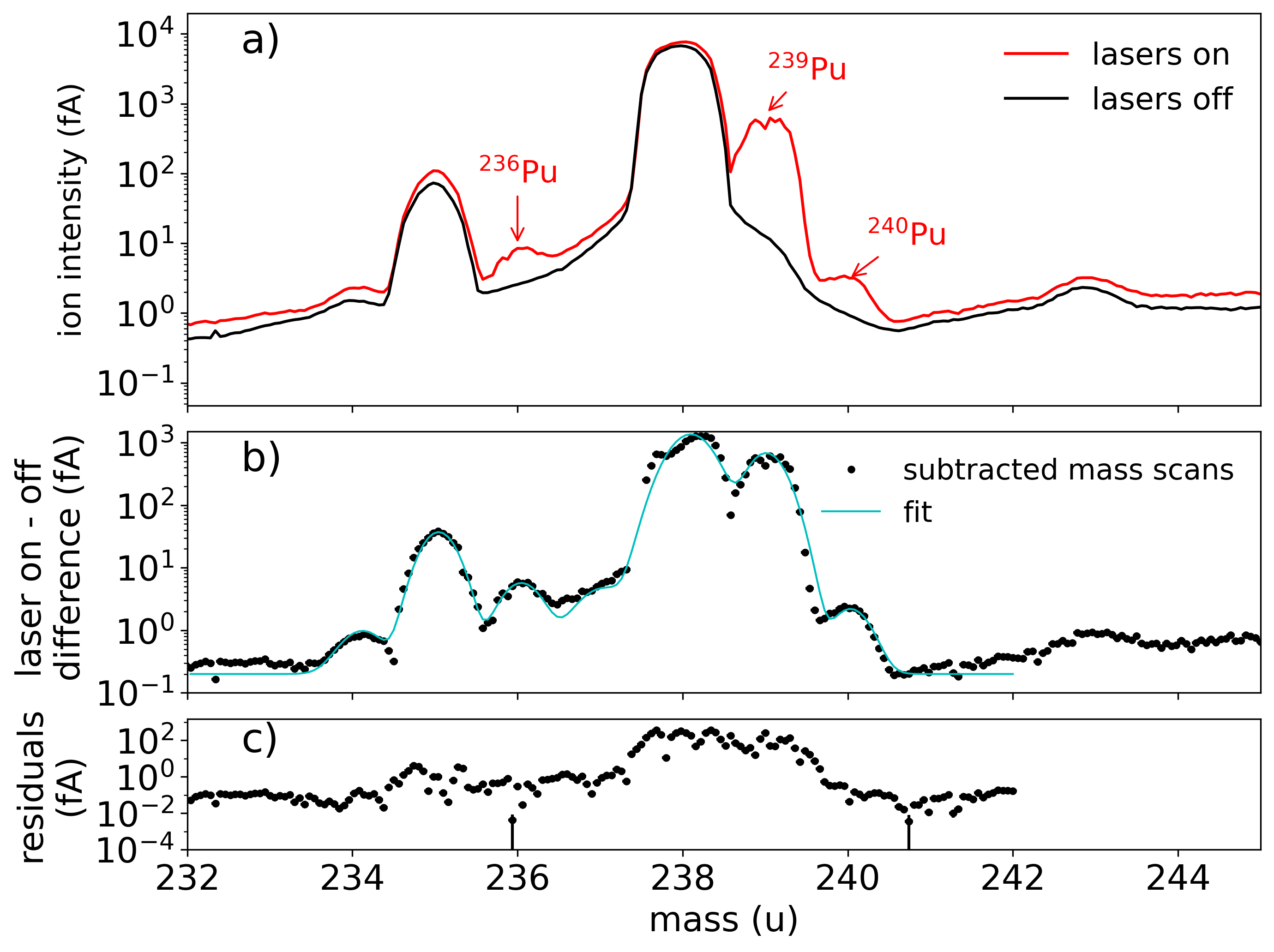}%
\caption{a) Signal on the MagneToF detector during a mass scan of the ISOLDE GPS  in the ``after, off" condition. The mass spectrum taken with the laser scheme for resonant ionization of Pu (Fig.~\ref{fig:schematic}) is shown in red. The mass spectrum with the FES laser off resonance is shown in black. b) The difference between ``lasers on" and ``lasers off" mass scans in a) fitted with a model of combined Gaussians. c) Difference between the experimental data and the fit shown in b). \label{fig:massscan_pu}}
\end{figure}
After irradiating the target with additional proton beam and heating to 1990\,$^{\circ}$C, a mass scan of the GPS separator magnet on the MagneToF detector showed a laser effect for Pu on the masses 236, 239 and 240 (Fig.~\ref{fig:massscan_pu}). The laser wavelengths were optimized on $^{239}$Pu. Fitting the difference between the mass spectra with lasers on and lasers off with a sum model of seven Gaussian peaks fixed at the masses of $^{234-240}$Pu (Fig.~\ref{fig:massscan_pu}) gives peaks corresponding to the laser response on each mass. For $^{237}$Pu, only an upper limit can be extracted. 

\begin{figure}
	\includegraphics[width=0.45\textwidth]{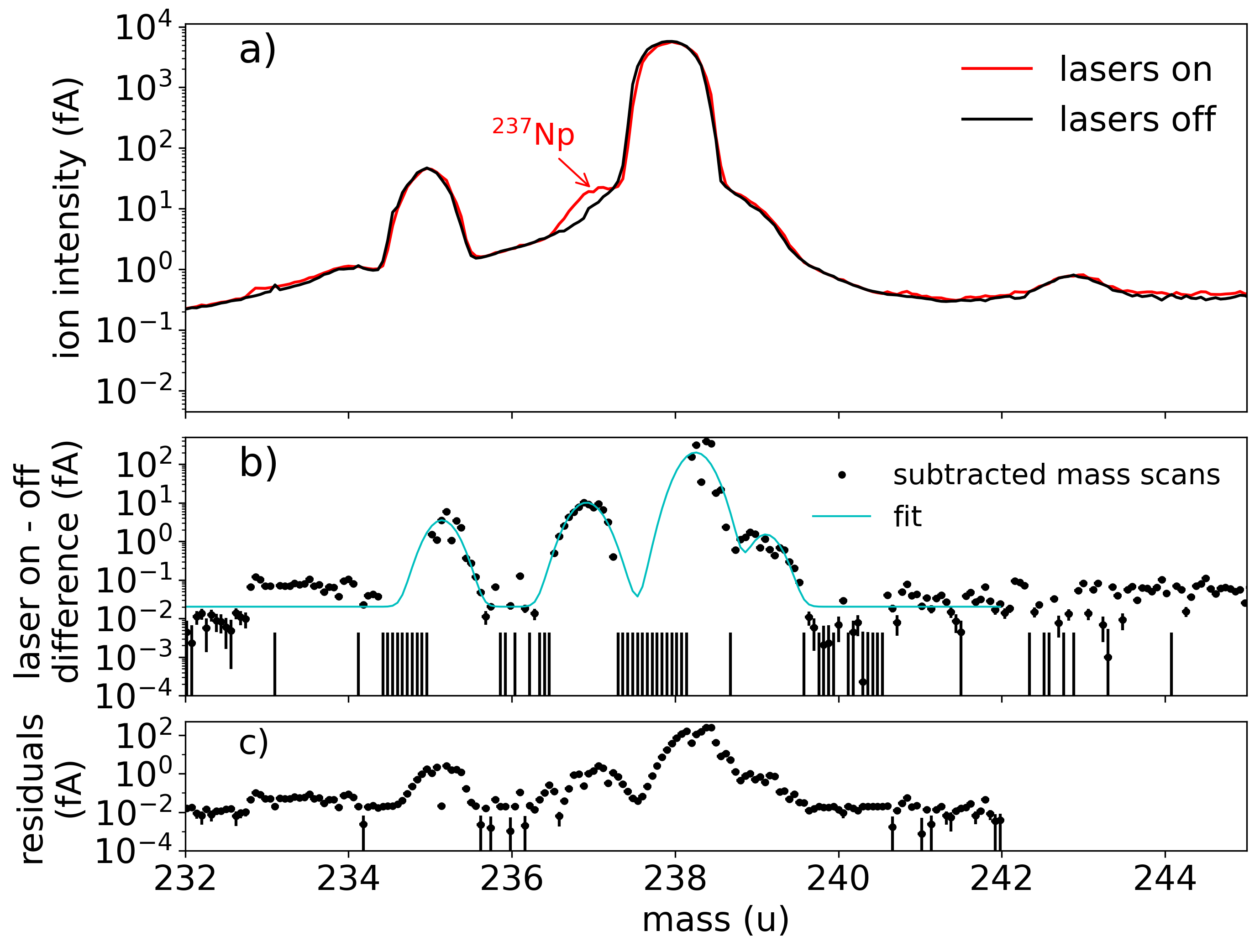}%
	\caption{a) Signal on the MagneToF detector during a mass scan of the ISOLDE GPS in the ``after, off" condition. The mass spectrum taken with the laser scheme for resonant ionization of Np (Fig.~\ref{fig:schematic}) is shown in red. The mass spectrum with the FES laser off resonance is shown in black. b) The difference between ``lasers on" and ``lasers off" mass scans in a) fitted with a model of combined Gaussians. c) Difference between the experimental data and the fit shown in b). \label{fig:massscan_np}}
\end{figure}

For Np, a very small laser enhancement effect was visible in a mass spectrum at a target temperature of 2110\,$^{\circ}$C (Fig.~\ref{fig:massscan_np}). $^{237}$Np, the longest-lived Np isotope, can be seen as a shoulder on the large $^{238}$U peak. The difference was modeled with Gaussian peaks fixed at the masses of $^{235, 236-239}$Np.

\begin{figure}
	\includegraphics[width=0.45\textwidth]{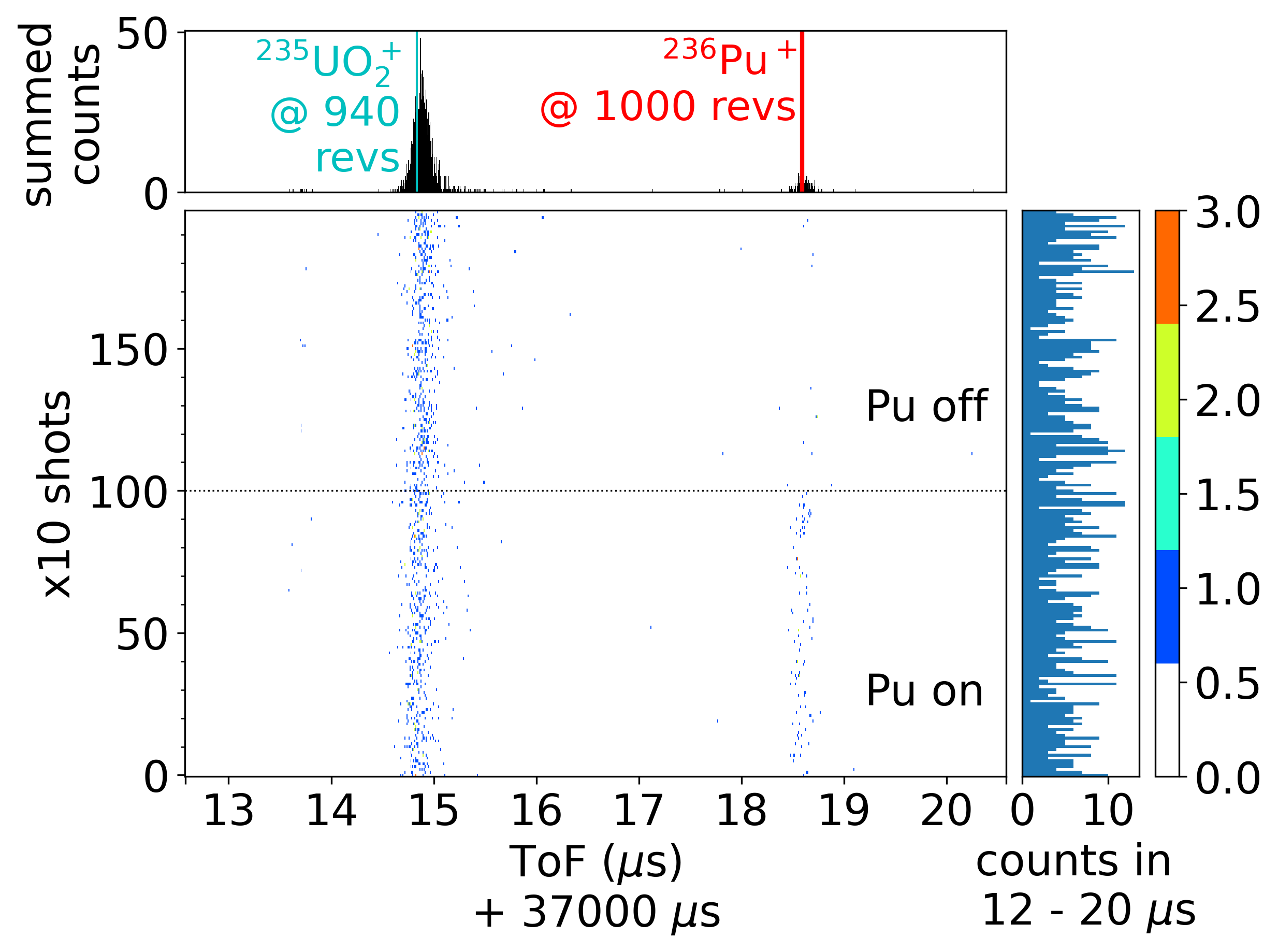}%
	\caption{ToF spectrum of mass 236 separated by the GPS with the Pu ionization scheme in the ion source. Trapping the ions for 1000 revs in the ISOLTRAP MR-ToF MS gives two visible ion ToF traces in the ToF range of interest for $^{236}$Pu (red). The Pu FES was blocked around 1000 shots, which led to a significant reduction in intensity of the trace around 37018.5\,$\mu$s. \label{fig:heatmap236}}
\end{figure}

\begin{figure}
	\includegraphics[width=0.5\textwidth]{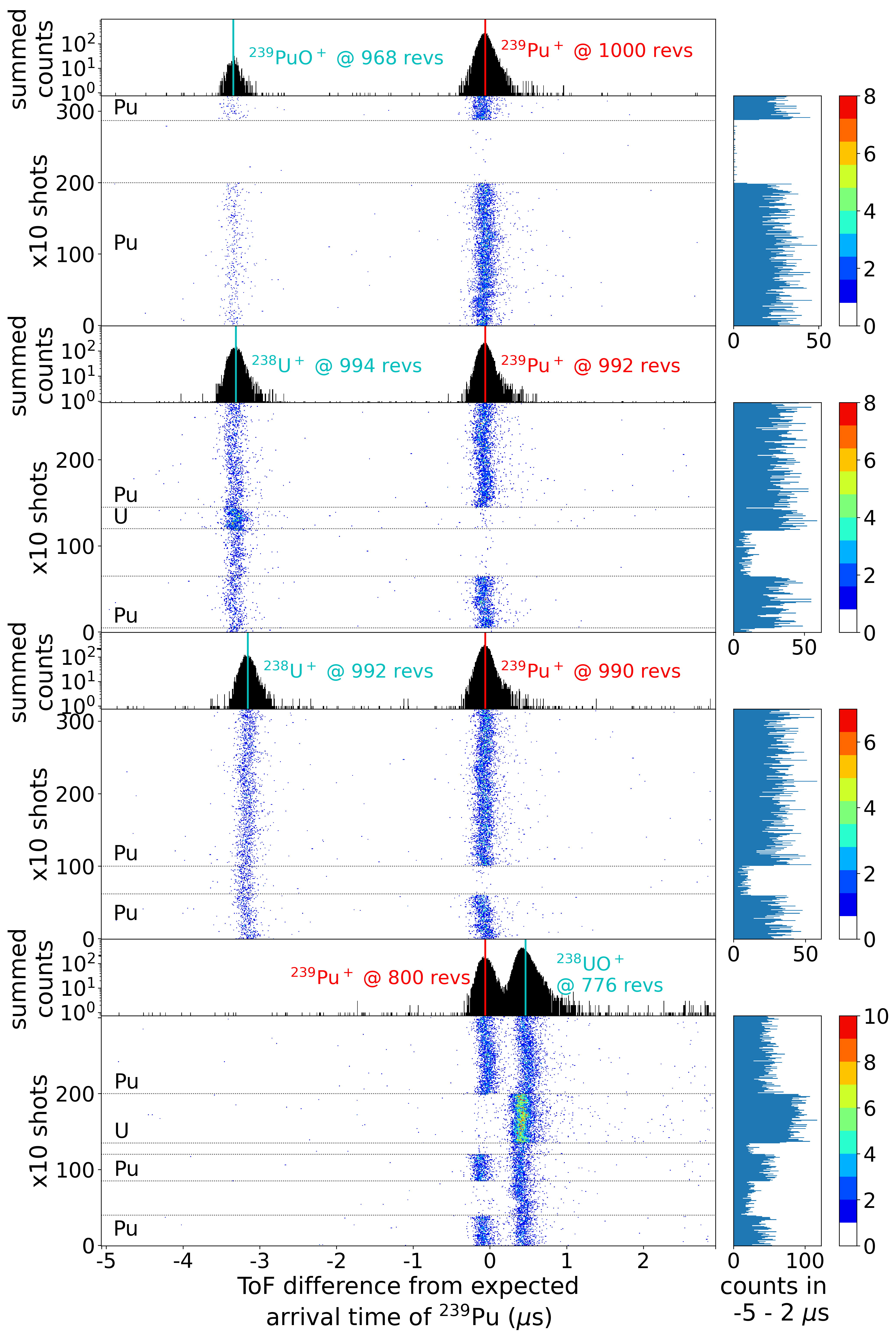}%
	\caption{ToF spectrum of mass 239 separated by the GPS, trapped for 1000, 992, 990 and 800 revolutions in the ISOLTRAP MR-ToF MS. Vertical axes show the summed counts and counts per 10 cycles (shots) for each number of revolutions. The color scale indicates the number of counts in a 10\,ns time bin. Horizontal axes show ToF since bunch ejection from the RFQ-cb with respect to the predicted ToF for $^{239}$Pu. Horizontal dotted lines indicate a change in the laser configuration; laser scheme(s) applied (U, Pu) are indicated on the left. \label{fig:heatmap239}}
\end{figure}

ToF spectra from the MR-ToF MS were used for beam composition identification. Non-isobaric contamination in an isobaric ToF spectrum can be matched with the corresponding mass on a given number of revolutions by using the ToF calibration (e.g. Fig.~\ref{fig:heatmap236}). In cases where a laser scheme was available for the contaminant (e.g. $^{235,238}$U from the target material), the laser effect on the contaminant was used for additional identification (see Fig.~\ref{fig:heatmap239}).

On nominal mass 236, $^{236}$Pu was identified in the MR-ToF MS after trapping for 1000 revolutions using laser response. It was present in the ToF spectrum along with a non-isobaric contaminant that showed no change in intensity with the Pu FES. The contaminant arrived 3.7\,$\mu$s earlier than the $^{236}$Pu. This corresponds within 60\,ns to the expected time-of-flight of $^{235}$U$^{16}$O$_2$ on 940 revolutions (Fig.~\ref{fig:heatmap236}). Since the mass resolving power of the GPS would have prevented simultaneous injection of Pu and UO$_2$ into the ISOLTRAP RFQ-cb, some surface-ionized $^{235}$U appears to have formed $^{235}$U$^{16}$O$_2$ in the RFQ-cb. $^{235}$U is also known to be present in the target material and is an expected contaminant that would not be fully suppressed by the separator magnet. 

On nominal mass 239, $^{239}$Pu was identified with the ISOLTRAP MR-ToF MS along with contaminants. At 1000 revolutions, the ToF centroid of the contaminant was earlier than the $^{239}$Pu as shown in Fig.~\ref{fig:heatmap239}. The contaminant at 1000 revolutions responded to the Pu laser scheme and matched the predicted time-of-flight for $^{239}$Pu$^{16}$O at a lower number of revolutions. Laser-ionized $^{239}$Pu appears to have formed $^{239}$Pu$^{16}$O in the RFQ-cb. With trapping time corresponding to 992 and 990 revolutions of $^{239}$Pu, the $^{239}$PuO was not visible, confirming that it is non-isobaric. $^{238}$U from the target material was identified as another contaminant. The ToF trace matched the expected ToF of $^{238}$U and responded to the U laser ionization scheme, but not the Pu laser scheme. On 800 revolutions another contaminant was identified, matching the expected ToF of $^{238}$U$^{16}$O. 

\begin{figure}
	\includegraphics[width=0.45\textwidth]{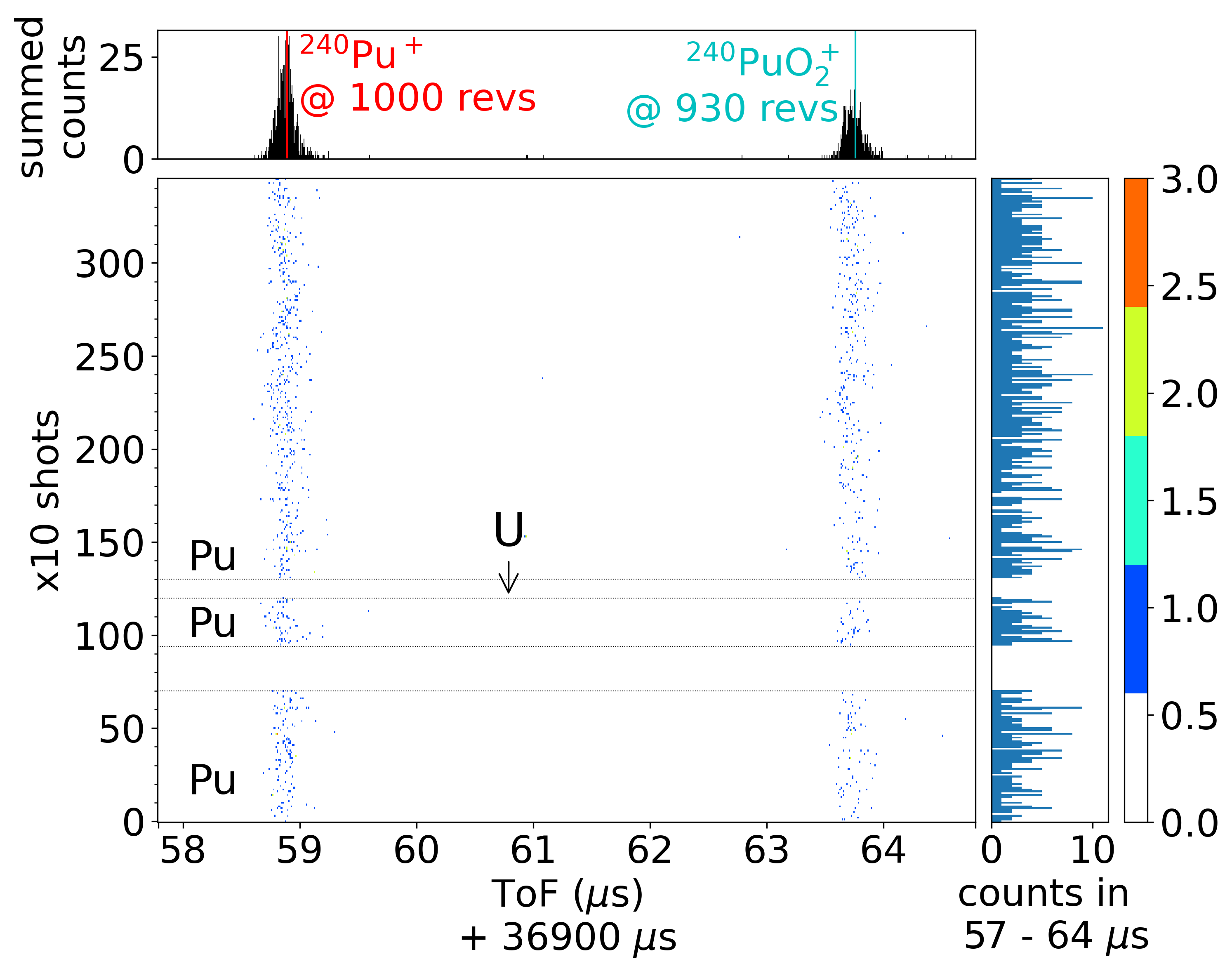}%
	\caption{ToF spectrum of mass 240 separated by the GPS and trapping for 990 revs in the ISOLTRAP MR-ToF MS. Red and blue show the expected ToFs for $^{240}$Pu and $^{240}$Pu$^{16}$O$_{2}$ at 990 and 930 revolutions, respectively. Horizontal dotted lines indicate a change in the laser configuration; laser scheme(s) applied (U, Pu) are indicated. \label{fig:IDs240}}
\end{figure}

On nominal mass 240, $^{240}$Pu was identified in the MR-ToF MS with only one trace visible at 1000 revolutions. On 990 revolutions, a contaminant was present 4.86\,$\mu$s later than $^{240}$Pu, which matched with the expected ToF of $^{240}$Pu$^{16}$O$_2$ on 930 revolutions (Fig.~\ref{fig:IDs240}). Both traces dropped noticeably in count rate when the Pu FES was blocked. No count rates above background were observed on this mass with the U laser in the ion source and the Pu FES blocked. 

\begin{figure}
	\includegraphics[width=0.5\textwidth]{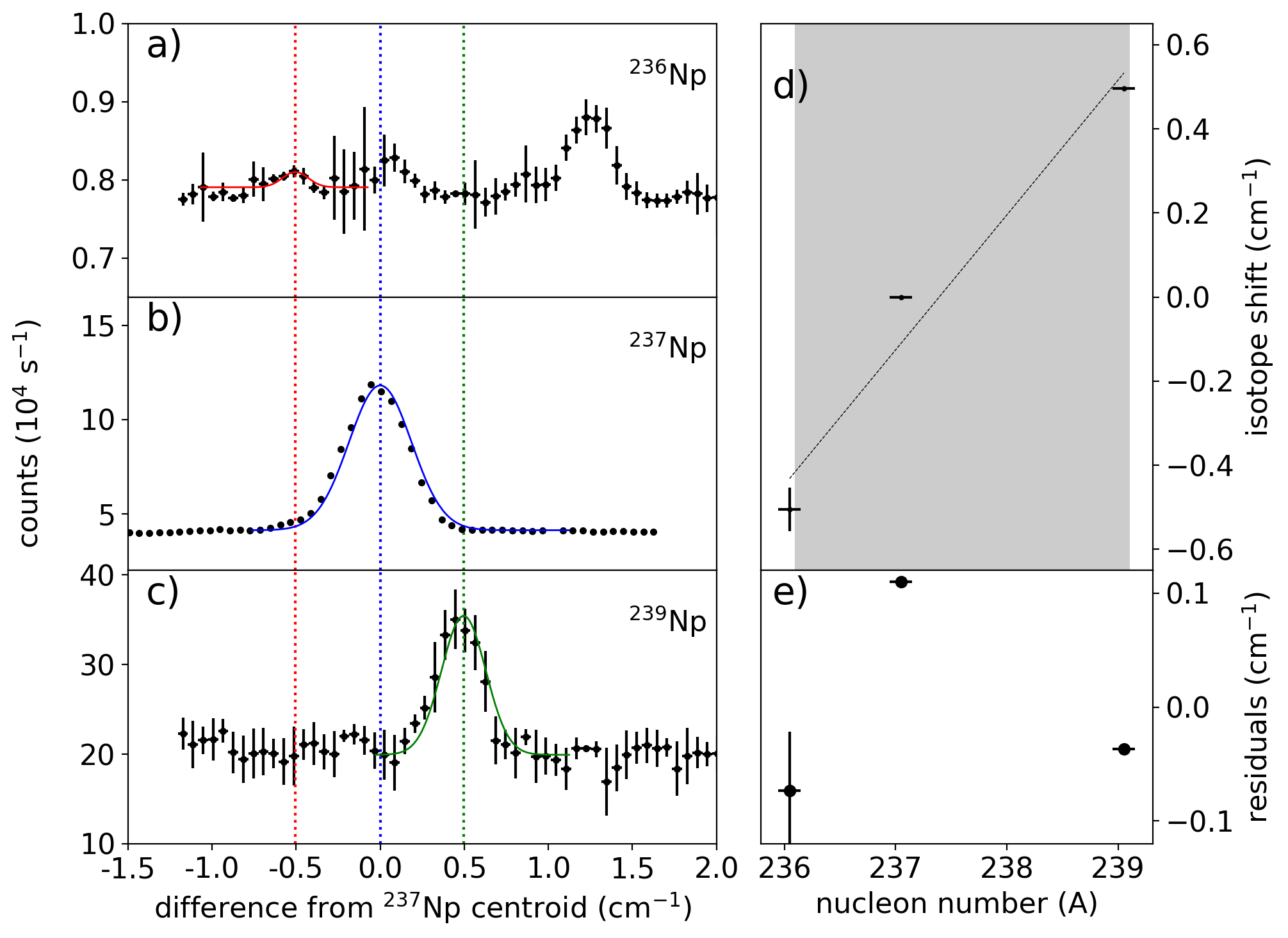}%
	\caption{Intensity on the MagneToF detector vs. the laser wavenumber of the Np FES for mass-separated ion beams: a) A=236, b) A=237, c) A=239 from the GPS. Wavenumbers are with respect to the centroid of the FES resonance of $^{237}$Np, shown as the dotted blue line in b). d) Plot of isotope shift with respect to $^{237}$Np and linear best fit (black dashed line). The fit's large 95\% confidence interval indicated in grey extends beyond the plot range. e) residuals shown as the difference between the measured isotope shift and the best fit shown in d). \label{fig:np_isotope_shifts}}
\end{figure}
\begin{figure}
	\includegraphics[width=0.5\textwidth]{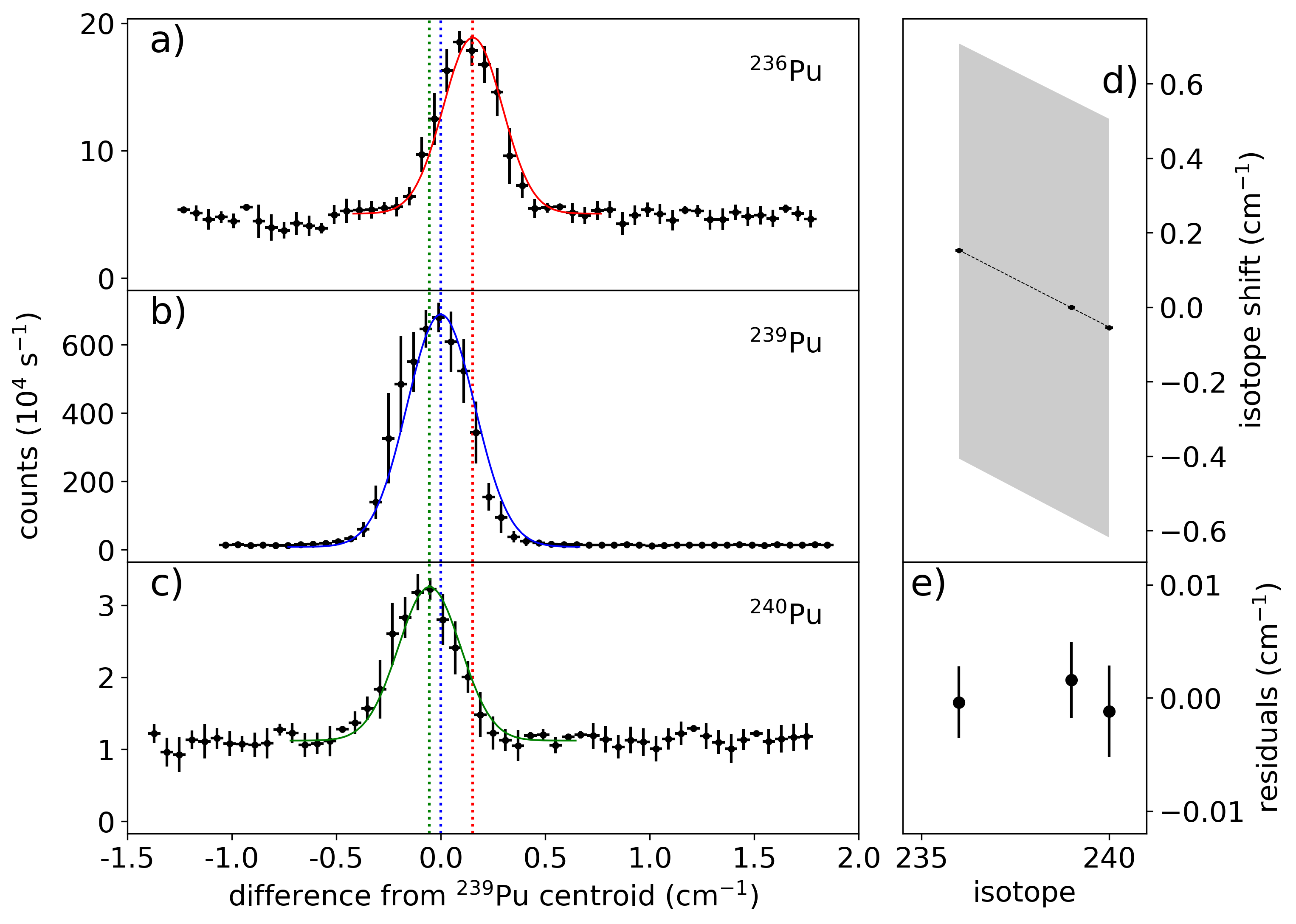}%
	\caption{Intensity on the GLM MagneToF detector vs. the wavenumber of doubled Ti:Sa light of the Pu FES for mass-separated ion beams: a) A=236, b) A=239, c) A=240 from the GPS. Wavenumbers in a,b,c) are with respect to the centroid of the FES resonance of $^{239}$Pu, shown as the dotted blue line in b). d) Plot of isotope shift with respect to $^{239}$Pu and linear best fit (black dashed line) with its 95\% confidence interval indicated in grey. e) residuals shown as the difference between the measured isotope shift and the best fit shown in d). \label{fig:pu_isotope_shifts}}
\end{figure}

The 395.6-nm transition in Np and the 413.4-nm transition in Pu were scanned for 3 isotopes each (Figs.~\ref{fig:np_isotope_shifts},\ref{fig:pu_isotope_shifts}) using the grating Ti:Sa laser. The Gaussian fit gave a FWHM of approximately 5\,GHz. From the three isotopes, the linear best fits found for the isotope shifts with respect to nuclide mass were: $y_{\textrm{Np}} = 0.320(63)m-76(15)$\,cm$^{-1}$ and $y_{\textrm{Pu}} = -0.0514(7)m+12.29(0.16)$\,cm$^{-1}$. For Np, deviations from linearity are larger than experimental uncertainty, suggesting that the isotope shifts are not linear. $^{236}$Np, with an odd neutron number, appears to exhibit a smaller isotope shift than the extrapolation of the data points $^{237,239}$Np, which both have even numbers of neutrons. For Pu, deviations from linearity are within the absolute uncertainty of the wavemeter (0.03\,cm$^{-1}$) and the uncertainty given by the linear best fit.

\begin{figure}
	\includegraphics[width=0.4\textwidth]{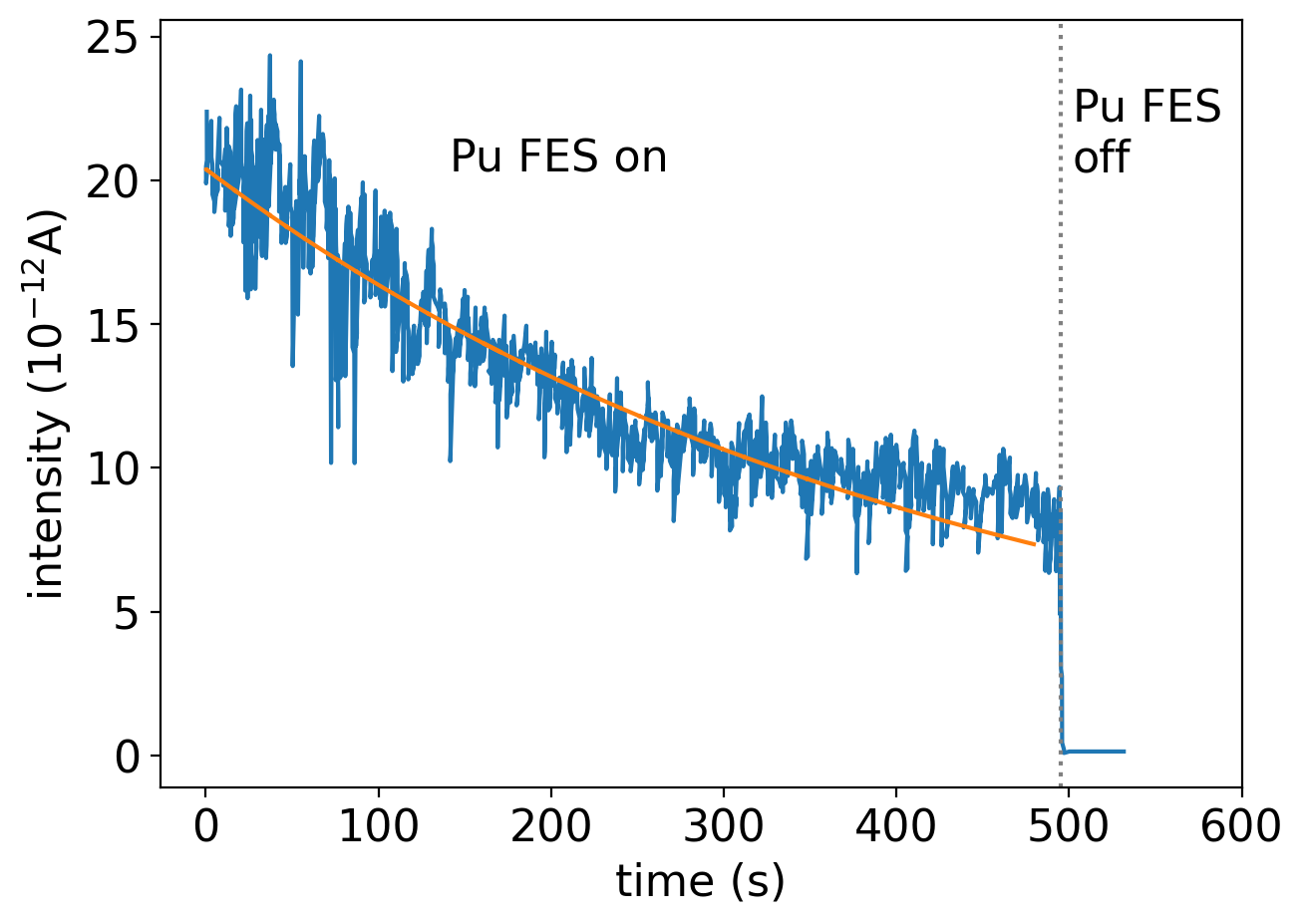}%
	\caption{Release response of $^{239}$Pu at 1990\,$^{\circ}$C target temperature after the target was irradiated with an integrated current of 1\,$\mu$A of protons. Data is shown in blue and the best fit using an exponential decay is shown in orange. Time 0 indicates the stop of irradiation. The grey vertical line indicates blocking the Pu FES from the ion source and the resulting surface ionized background. \label{fig:release_239}}
\end{figure}

The time structure of isotope release from thick targets exhibits an element-dependent behavior that limits extraction of short-lived isotopes using the ISOL technique. For $^{239}$Pu, the release time was evaluated by recording the response to stopping direct proton irradiation (Fig.~\ref{fig:release_239}) after irradiation with a proton intensity of 1\,$\mu$A with a constant ion current indicating steady-state conditions of production and release. For Pu, steady-state conditions were achieved after irradiation times on the order of 600\,s. The signal on an FC for $^{239}$Pu was recorded immediately after stopping irradiation. An exponential of the form $A \exp{\frac{-t}{\tau}}+C$ was used to fit the data. Here, $C$ represents the constant background on the FC. The amplitude $A$ of the exponentially falling signal indicates the rate of isotopes generated by protons on target. The exponential decrease captures the combined release following many individual pulses, and therefore sets an upper limit for the time required to extract a Pu isotope from the UC$_\mathrm{x}$ target. At a target temperature of 1990(50)\,$^{\circ}$C, this upper limit of the extraction time constant $\tau$ was evaluated to be 430(34)\,s. For the 24110-year half-life of the $^{239}$Pu used in this experiment, the contribution of radioactive decay to the decrease of signal during the 500\,s measurement is 6.6$\times10^{-10}$ of the initial amount and is therefore negligible in the determination of the release time. 

\begin{figure}
	\includegraphics[width=0.5\textwidth]{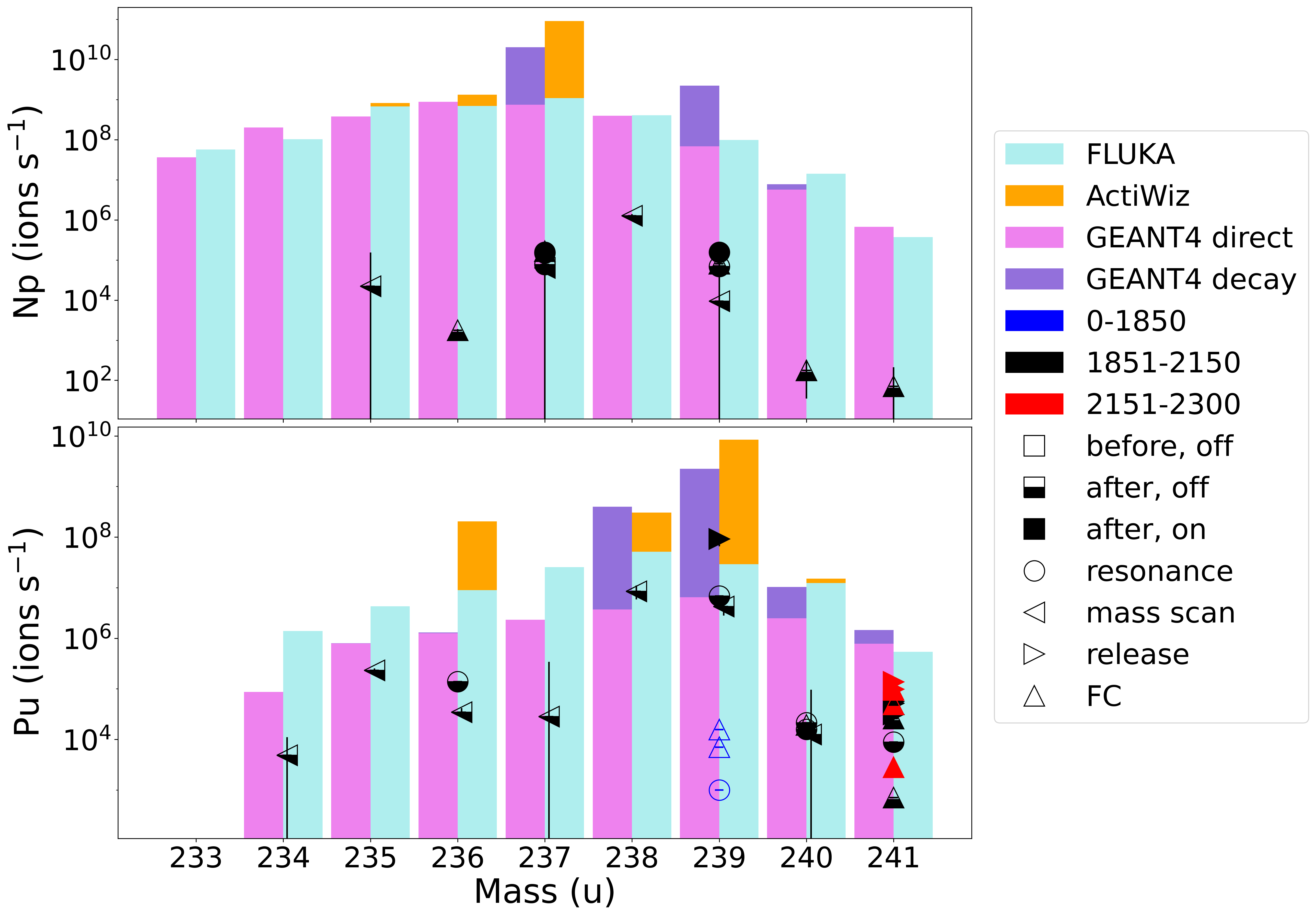}%
	\caption{Combined rates of laser-ionized Np (top) and Pu (bottom) beams from laser resonances (circles), mass scans (left arrows), protons on/off measurements (right arrows), and FC measurements (up arrows) shown with predicted total in-target production rates from the FLUKA and G\textsc{eant}4 models as bars. The rates from direct reactions are shown in light blue and pink for FLUKA and G\textsc{eant}4 respectively. The fractions from decay reactions are shown using orange and purple bars for ActiWiz and G\textsc{eant}4 respectively. Marker colors indicate the target temperature in $^{\circ}$C during the measurement (blue: below nominal, black: nominal, red: above nominal) and the marker fill indicates the irradiation state during the measurement (empty: offline measurement before re-irradiation, half-filled: measurement taken after re-irradiation with the proton beam ``off", filled: online measurement during irradiation).  \label{fig:rates}}
\end{figure}

The modeled inventory generated by 1\,$\mu$C of proton primaries using G\textsc{eant}4 and FLUKA is compared against the measured values in Fig.~\ref{fig:rates}. Resonance signal height on each mass and peak heights from the sum of the Gaussian fits gives a rate of the laser-ionized species without considering the surface-ionized species or surface-ionized contaminants. ToF compositions indicate negligible contributions from surface ionization of Pu compared to the laser-ionized Pu ions. From evaluation of the isotope shifts and the laser linewidth of approximately 5 GHz, the effect in intensity of the different isotopes can be estimated (Figs.~\ref{fig:np_isotope_shifts},\ref{fig:pu_isotope_shifts}).

\begin{table}
	\caption{Maximum rates reported for Np and Pu ions at target temperatures between 1850 and 2150\,$^{\circ}$C. Measurements where uncertainties are larger than the measured value are not included. \label{tab:rates}}
	\begin{ruledtabular}
		\begin{tabular}{c c c}
			mass (u) & Np (ions/s) & Pu (ions/s)\\
			\hline
			235 & 2(1)E+4 & 2.3(2)E+5\\
			236 & 1.76(14)E+3 & 1.38(2)E+5	\\
			237 & 1.56(51)E+5 & $<$ 2.8E+4\\
			238 & 1.3(1)E+6 & 8.5(2.6)E+6\\
			239 & 1.55(4)E+5 & 6.82(9)E+6\\
			240 & 1.7(14)E+2 & 2.14(4)E+4\\ 
			241 & 7.1(4)E+1 & 2.6(7)E+4\\
		\end{tabular}
	\end{ruledtabular}
\end{table}

\section{Discussion} \label{discussion}
Isotope shifts are connected to the changes in nuclear charge radii and are therefore not necessarily expected to be linear with neutron number \cite{Yang2023}, especially in the actinide region of the nuclear chart \cite{Block2021}. The $^{236,239,240}$Pu isotope shifts agree with linearity, but the $^{236,237,239}$Np isotope shifts are not well-described by a linear fit. The offsets could suggest odd-even staggering or the presence of a kink in the nuclear charge radius. These preliminary measurements motivate future campaigns to extract further information about the nuclear charge radii of Np.

From laser resonances, laser on/off mass scans and proton on/off measurements, ion beam rates are reported for $^{235-241}$Pu (Fig.~\ref{fig:rates}). These rates are achievable with the use of pre-irradiated uranium carbide targets combined with production of isotopes through inelastic reactions during irradiation and can be achieved at nominal (1850-2150\,$^{\circ}$C) UC$_\textrm{x}$ target temperatures. In contrast, rates of Np isotopes are below $10^6$\,ions\,s$^{-1}$ even with protons on target and target temperatures above 2100\,$^{\circ}$C, indicating poor release efficiency as follows from a comparison with simulated production rates. Maximum rates observed experimentally in this work are reported in Tab.~\ref{tab:rates}. Further developments may be required to enhance extraction from the target or improve the efficiency of resonance ionization laser schemes. 

Mass scans and ToF measurements of ion beams mass-separated by the separator magnets show considerable rates of surface-ionized $^{235,238}$U on the nearby masses of interest. The mass-separated actinide beams from ISOLDE formed molecules in the ISOLTRAP RFQ-cb, creating contaminants several mass units away from the species of interest, particularly for U and Pu. These contaminants account for non-isobaric contaminants in the ToF spectra. In many cases the ToF could be matched with a laser effect and corresponding number of revolutions to identify a likely contaminant (Fig.~\ref{fig:heatmap239}). The molecular formation decreased the effective efficiency of the species of interest, therefore ion rates from the MR-ToF MS measurements are not reported.  

The G\textsc{eant}4 model (Table \ref{tab:G4production}) predicts trends in production mechanisms for these above-target isotopes. The light Np isotopes $^{228-236}$Np are produced to more than 99\% through inelastic reactions, with direct irradiation contributing more than 85\% to the total production for all of these isotopes. $^{237,239,240}$Np show large production fractions through decay, with 96.3\%, 97.0\% and 27.0\%, respectively. $^{240, 241}$Np were mostly produced through inelastic reactions with $^{240}$Np mainly from tritons (47.8\%) and $\alpha$-particles (18.8\%), and $^{241}$Np dominantly from $\alpha$-particles (96.3\%) and heavier ions (3.7\%). Heavier isotopes $^{242-244}$Np were modeled with less than 5 events each, entirely from heavy ions. For Pu, inelastic reactions with protons and $\alpha$-particles contributed dominantly to the modeled production of the lighter isotopes $^{234-237}$Pu, while the heavier isotopes $^{238-244}$Pu result largely from the $\beta$-decay of Np isotopes and, for isotopes above $^{240}$Pu, also from heavy ion inelastic reactions. For $^{237,239}$Np and $^{238,239}$Pu, decay contributions are described in both models. The FLUKA model and the direct reaction contribution from the G\textsc{eant}4 model agree well for Np but show slightly different behaviour in the case of Pu. Differences in the predictions of decay contributions given by the ActiWiz model and the G\textsc{eant}4 model could be attributed to the difference in the modeled cooling time.

Events of $^{235-240, 242-244}$Am were modeled to be between 1 and 100 events per 10$^9$ primary protons. 162 events of $^{241}$Am were modeled dominantly from $^{241}$Pu decay. Two of the Am isotopes seen in the model, $^{242,244}$Am, have short half lives (16\,h and 10\,h respectively) and $\beta$-decay into $^{242,244}$Cm. Neither $^{242}$Cm nor $^{244}$Cm decay into the next element, Bk, marking the $Z$-limit of elements available for experiments at high-energy proton-accelerator-based thick target facilities. To reach usable rates of nuclides with higher proton numbers, these facilities could employ heavier target nuclei, improve extraction efficiency by several orders of magnitude, or improve experimental sensitivity and increase duration. There are limited practical options for targets of bulk material with nuclei heavier than $^{238}$U, where decreasing half-lives and available quantities of actinide targets come with increased costs and radioactivity handling considerations. These results demonstrate the availability of two new transuranium beams from UC$_\textrm{x}$ targets at the ISOLDE facility with rates typically required for current experimental sensitivity. Additionally, comparisons of production pathways with measured rates suggest the potential availability of a select few Am isotopes for experiments with state-of-the art sensitivity.

\section{Conclusions}

From 1.4-GeV proton irradiations of a UC$_x$ target, above-target nuclides of neptunium and plutonium have been produced, released from the hot target, resonantly laser-ionized and identified as ion beams at the ISOLDE facility. Production mechanisms modeled by G\textsc{eant4} predict production-pathway-specific yields. On the neutron-deficient side of stability, Np and Pu isotopes are formed through inelastic reactions, mostly induced by protons. Decay contributions from $\beta$-decaying U precursors are significant for $^{237,239,240}$Np and $^{238-244}$Pu, with the $\beta$-decay of Np contributing to the Pu inventory. The use of previously irradiated targets with long cooling times thus contributes to the in-target inventory especially of Pu isotopes, while neutron-deficient isotopes are produced dominantly from direct irradiation. 

Two-step ionization schemes using intra-cavity doubled Ti:Sa lasers were applied to resonantly ionize both Np and Pu. Isotope shifts were investigated for the 395.6-nm ground state transition in $^{236,237,239}$Np and the 413.4-nm ground state transition in $^{236,239,240}$Pu. To our knowledge, there is no data available in literature for the charge radius of Np nuclei. With two odd-odd nuclei and odd-even $^{236}$Np, Np isotope shifts did not follow a linear trend. Further measurements are required to extract information about the nuclear charge radii. In the case of Pu, isotope shifts are linear with nucleon number in the investigated range and can be extrapolated for use over a wider range of isotopes. 

Ion production rates were measured for $^{235-240}$Np and $^{234-240}$Pu (Fig.~\ref{fig:rates}). ToF MS spectra show surface-ionized contamination on the masses of interest, originating from the uranium carbide target material. Oxide formation is observed in a gas-filled RFQ-cb. Comparing the observed rates with predicted in-target inventory gives an estimation for the extraction efficiency of Np on the order of 0.001\% and Pu on the order of 0.1\%. Target temperatures required for observable rates of Np were above 2100\,$^{\circ}$C, while target temperatures required for Pu were above 1150\,$^{\circ}$C. Pu is released within 400-500\,s at nominal target temperatures. The demonstrated availability of Np and Pu beams at ISOLDE brings two new actinide elements into reach for experiments at ISOL facilities with useful intensities of specific isotopes as functions of different experimental parameters such as decay time. Information on laser ionization schemes, magnitudes of isotope shifts, and release temperatures informs future experiments requesting beams of these elements. 

$^{241}$Pu is the heaviest nuclide identified at a proton-accelerator-driven ISOL facility to date. While extremely sensitive experiments may be capable of extending studies one element higher to specific isotopes of Am discussed here, the investigation of production mechanisms suggests that plutonium is the high-$Z$ limit of heavy nuclide production from 1.4-GeV protons on a $^{238}$U target with typical rates required for current experimental sensitivity.

\begin{acknowledgments}
The authors gratefully acknowledge technical support from E. Barbero, M. Bovigny, B. Crepieux, J. Cruikshank and the ISOLDE operations staff. This project has received funding from the European Union's Horizon 2020 Research and Innovation Program (grant No. 861198 project ‘LISA’ MSCA ITN). The authors acknowledge support from the German Federal Ministry of Education and Research (BMBF) for ISOLTRAP (grant No. 05P18HGCIA and 05P21HGCI1). L.N. acknowledges support from the Wolfgang Gentner Program (grant No. 13E18CHA).
\end{acknowledgments}

\appendix
\section{In-target production models}
\label{appendix:models}
The FLUKA Monte-Carlo model of in-target production is a standard tool available for users of the CERN-ISOLDE facility, with a database of modeled values \cite{Ballof2020}. This model uses the ABLA description of nuclear de-excitation through particle evaporation and fission for the decay pathways of nuclear systems \cite{Kelic2009}. The FLUKA model incorporates no cooling time for radioactive decay. Instead, the FLUKA model of particle fluence spectra in the target was used as input to the ActiWiz Creator software version 3.4 \cite{Vincke2014}. ActiWiz is based on 100 CPU years of FLUKA calculations modeling interactions of protons, charged pions, photons, and neutrons above 20\,MeV with the evaluated libraries JEFF 3.3, ENDF VIII.0 and EAF2010 for neutrons below 20\,MeV. The ActiWiz model was used to extend the FLUKA model to include the inventory of radionuclides produced during the cooling period and is detailed in Ref.~\cite{Duchemin2020}.

G\textsc{eant}4 (version geant4-11) was used with physics list QGSP\_INCLXX\_HP\_ABLA, an experimental physics list using the Li{\`e}ge intranuclear cascade model for nucleon-nucleus interactions below 3\,GeV, with neutronHP (high precision neutron package) for neutron cross-sections below 20\,MeV and de-excitation modeled using ABLA with datasets for neutron, proton and pion cross-sections for elastic, inelastic, capture and fission reactions as described in \cite{G4physicsguide}. 

Predictions from the INCLXX and ABLA models show good agreement for a selection of lighter isotopic chains measured at ISOLDE \cite{Lukic2006}.

\section{Target and resonance ionization laser ion source}
\label{appendix:target}
The measured temperature was calibrated against the target and ion source heating current and required power using an optical pyrometer up to 2000\,$^{\circ}$C. The calibration data were extrapolated to estimate the target and ion source temperatures during the experiment with an uncertainty estimated to be 100\,$^{\circ}$C. The calibration was performed before the loading of the target material and the first irradiation of the target. 

Three diode pumped solid-state (DPSS) Nd:YAG lasers were used to pump a total of four titanium:sapphire (Ti:Sa) lasers based on wavelength selection by grating tuning \cite{Sonnenschein2015} or a combination of a Lyot filter and Fabry-Perot etalon \cite{Rothe2011} with a repetition rate of 10\,kHz. One intra-cavity—second-harmonic generation (IC-doubled) grating Ti:Sa laser \cite{Teigelhofer2010} was used to provide the first excitation step (FES) for either Pu or Np, with typical output power up to 400\,mW. The other three IC-doubled Ti:Sa lasers were used to provide the second excitation step (SES) of Pu, the identical wavelength of the FES and SES of U, and the SES of Np, each with up to 2.5\,W of output power. The Pu resonance ionization scheme was developed by Kneip et al \cite{Kneip2020} and uses a FES from the ground state to the $5f^{5}6d^{2}7s$ J=1 state followed by a SES to an auto-ionizing state located above the ionization potential. The U scheme was developed by Savina et al \cite{Savina2017} and features a two-step, single wavelength ionization from the ground state to an auto-ionizing state. Several schemes were investigated for Np using a FES of 25277.6\,cm$^{-1}$ from the ground state to a J=9/2 state.

To switch between ionizing Pu or Np, the grating Ti:Sa was changed between the Pu FES and the Np FES. Wavelength scans of the FES were done by changing the diffraction grating angle. Small-range wavelength scans of the second step were done manually by changing the etalon angle. The frequency of the Ti:Sa lasers was measured using a High Finesse WS7 wavemeter (uncertainty 20\% of the laser linewidth) to sample the fundamental, non-frequency-doubled mode. Atomic transitions in the ion source are expected to be Doppler-broadened, with the FWHM given by $\Delta \nu_{D}$ where: 
\begin{equation}
	\Delta \nu_{D} = \nu_{0}\sqrt{\frac{8k_{B}T\ln2}{mc^2}}	
\end{equation}

where $T$ is the absolute temperature, $k_{B}$ is the Boltzmann constant, $m$ is the mass of the atom, $\nu_{0}$ is the nominal laser frequency, and $c$ is the speed of light in vacuum. For the employed transitions around 400\,nm, the Doppler broadening at 2000\,$^{\circ}$C thus corresponds to Gaussian peak shapes with FWHM above 1.6\,GHz for a mass of 238\,u. Laser linewidth and power-broadening may increase the experimentally observed peak widths further \cite{Fedosseev2017}. 

\section{Ion beam analysis}
\label{appendix:beam}
The ion beams mass-separated by the GPS separator magnet were injected into the ISOLTRAP RFQ-cb, where they were collisionally cooled with the He buffer gas at pressures up to 10$^{-5}$\,mbar measured within 1\,m of the injection. After ejection from the RFQ-cb, the ion bunch energy was reduced to 3.2\,keV using a pulsed drift tube. Ion bunches were then injected into the ISOLTRAP MR-ToF MS \cite{Wolf2013} and trapped between the electrostatic mirror potentials using the in-trap lift method \cite{Wienholtz2013} for typically 1000 revolutions before release and detection of the ions' time of flight (ToF). The difference in ToF allows components with different mass-over-charge ratios $m/q$ to be separated in time $t$ with a mass resolving power $R$ given by:
\begin{equation}
	R=\frac{m}{\Delta m}=\frac{t}{2\Delta t}
\end{equation}
\vspace{1 mm}
where $t$ is the absolute time-of-flight and $\Delta t$ is the FWHM of the ToF distribution. After trapping times of up to 40\,ms, the ions were ejected onto the ISOLTRAP MagneToF detector. Ion arrival times were measured from the time of ejection from the RFQ-cb and recorded with 100\,ps resolution. Mass resolving powers $R$ in excess of 10$^5$ were achieved for the investigated mass range.

Additionally, non-isobaric beam components formed in the trapping and cooling process from the mass-separated beam can be identified. These components can appear in the same ToF spectrum when travelling through the in-trap lift cavity of the MR-ToF MS with the isobaric beam component on a different number of revolutions during bunch ejection. Non-isobaric components were identified by varying the trapping time of the isobaric beam bunch in the MR-ToF MS. This leads to different absolute ejection times, which increases the chance that the non-isobaric component is separated from the isobaric beam bunch. The ISOLTRAP MR-ToF MS was calibrated using $^{85,87}$Rb and $^{133}$Cs from the ISOLTRAP offline ion source and using $^{238}$U from the ISOLDE GPS target and ion source. The reference measurements were used to create a calibration to relate known masses to an expected ToF in the form: 
\begin{equation}
	t_{\mathrm{calibration}}(m) = a\sqrt{m/q} + b
\end{equation}
 where $a$ and $b$ are the calibrated parameters.

\bibliography{bib.bib,miscbib.bib}

\end{document}